%% file: main.tex
\documentclass{article} 
\usepackage{styles/iclr2020_conference,times}

\input{styles/math_commands.tex}

\input{styles/custom_commands.tex}

\usepackage{hyperref}
\usepackage{url}

\usepackage{listings}
\title{Privacy-preserving collaborative machine learning on genomic data using TensorFlow}

\author{Cheng Hong, Zhicong Huang, Wen-jie Lu, Hunter Qu\\
Gemini Lab, Alibaba Group \\
\texttt{\{vince.hc,zhicong.hzc,juhou.lwj,fuping.qfp\}@alibaba-inc.com} \\
\And
{Li Ma}\\
Alibaba Health \\
\texttt{ml96386@alibaba-inc.com}
\And
{Morten Dahl, Jason Mancuso}\\
Dropout Labs \\
\texttt{mortendahlcs@gmail.com,jason@manc.us}
}

\iclrfinalcopy 

\begin{document}


\maketitle
 
\begin{abstract}
Machine learning (ML) methods have been widely used in genomic studies. However, genomic data are often held by different stakeholders (e.g. hospitals, universities, and healthcare companies) who consider the data as sensitive information, even though they desire to collaborate. To address this issue, recent works have proposed solutions using Secure Multi-party Computation (MPC), which train on the decentralized data in a way that the participants could learn nothing from each other beyond the final trained model. 

We design and implement several MPC-friendly ML primitives, including class weight adjustment and parallelizable approximation of activation function. In addition, we develop the solution as an extension to TF Encrypted~\citep{dahl2018private}, enabling us to quickly experiment with enhancements of both machine learning techniques and cryptographic protocols while leveraging the advantages of TensorFlow's optimizations. Our implementation compares favorably with state-of-the-art methods, winning first place in Track IV of the iDASH2019 secure genome analysis competition. \footnote{http://www.humangenomeprivacy.org/2019/}
\end{abstract}

\input{1_Intro.tex}

\input{2_Background.tex}
\input{3_Solution.tex}
\input{4_Results.tex}
The final submission was tested in a LAN network with slightly higher latency. We achieved the highest accuracy ($68-70\%$) and the second fastest training speed (560 iterations in about 20 seconds) among the nine participating teams, and were named first place in a tie of three.
\section*{Acknowledgements}
We thank Peter Rindal from Visa Research for helpful discussion on improving Logistic Regression implementations in ABY3.

\input{Appendix.tex}

\bibliographystyle{styles/iclr2020_conference}
\bibliography{bibs/privacy}

\end{document}

%% file: styles/math_commands.tex
\usepackage{amsmath,amsfonts,bm}
\usepackage{algorithm}

\usepackage{amssymb}
\usepackage[noend]{algpseudocode}

\def\eqref#1{equation~\ref{#1}}

\def\1{\bm{1}}

\DeclareMathAlphabet{\mathsfit}{\encodingdefault}{\sfdefault}{m}{sl}
\SetMathAlphabet{\mathsfit}{bold}{\encodingdefault}{\sfdefault}{bx}{n}

\DeclareMathOperator*{\argmin}{arg\,min}

%% file: styles/custom_commands.tex
\newcommand{\mypara}[1]{\vspace*{0.05in}\noindent\textbf{#1.}$\;$}

\usepackage{multirow}
\usepackage{graphicx}
\usepackage{booktabs}

%% file: 1_Intro.tex
\section{Introduction}

Machine learning methods have been applied to a huge variety of problems in genomics and genetics \citep{libbrecht2015machine}. A typical example is to train a model to classify healthy and (potentially) diseased people according to their genomic information. Generally speaking, larger amount of training data is required to make more successful ML models. Unfortunately, genomic data are considered to be highly sensitive information for individuals, and thus are usually held by different data owners in a strictly access-controlled way \citep{cho2018secure}. Therefore, it becomes highly important to allow two or more genomic data owners to jointly train a model without compromising each other's data privacy. 

iDASH (integrating Data for Analysis, Anonymization, Sharing), a National Center for biomedical computing funded by National Institutes of Health (NIH), has hosted a secure genome analysis competition for the past 5 years. This contest has encouraged cryptography experts all over the world to develop secure and practical solutions for privacy-preserving genomic data analysis. Specifically, the iDASH competition announced four tracks this year, with Track IV calling for solutions on secure collaborative training of ML model using MPC. The organizers provided two genomic datasets: GSE2034, containing 142 positive and 83 negative tumor samples with 12,634 features each, and BC-TCGA, containing 422 positive and 48 negative tumor samples with 17,814
features each.

The challenge of Track IV is threefold: \textbf{1)} A protocol in the 3-party semi-honest (with honest majority) model is required, but implementations of recent state-of-the-art MPC protocols for solving such machine learning tasks were not available\footnote{The code of \cite{aby3} came out in 2019.7, two months after the competition began.}.  \textbf{2)} It is hard to avoid overfitting due to the small sample size and large number of features. \textbf{3)} The dataset is heavily imbalanced but common countermeasures such as resampling are difficult in MPC. 

We summarize our contributions as follows: 

\begin{itemize}

\item We implemented the state-of-the-art \textbf{ABY3} \citep{mohassel2018aby} protocol using the TF Encrypted framework~\citep{dahl2018private}. With the advantages of TF Encrypted, our implementation is $1.1-1.8\times$ faster than the original ABY3 implementation for large-scale ML training. The code has been open-sourced to the TF Encrypted repository.

\item We developed a secure collaborative ML solution on top of our TF Encrypted-ABY3 framework, together with several MPC-friendly ML primitives, including class weight adjustment for the imbalanced dataset and more accurate and parallelized sigmoid approximation. The solution tied for first place in Track IV of iDASH2019 competition.

\end{itemize}

%% file: 2_Background.tex
\section{Background And Related Work}
\label{background}

\subsection{Logistic Regression}
The datasets have many more features than samples and are prone to overfitting, so we decided to use a simple logistic regression (LR) instead of more complex modeling approaches after initial experimentation. Let $\mathbf{x}\in \mathbb{R}^f$ denote the $f$-dimensional feature vector, $\mathbf{w}\in \mathbb{R}^f$ the corresponding weight vector, and $y\in\{0,1\}$ the corresponding label of $\mathbf{x}$, the goal of LR training is to solve the following empirical risk minimization problem:

\begin{center}
$\underset{\mathbf{w} \in \mathbb{R}^f}{\argmin} \, L(\mathbf{w}) = \underset{\mathbf{w} \in \mathbb{R}^f}{\argmin} \;  \underset{\mathbf{x}, y}{\mathbb{E}}
\bigg [y\log{\big(\sigma(\mathbf{w}^\top\mathbf{x})\big)} +(1-y)\log\big(1-\sigma(\mathbf{w}^\top\mathbf{x})\big)\bigg]$
\end{center}

While second-order Newtonian optimization is more commonly used in cleartext LR training, such methods are costly in MPC. We can instead use stochastic gradient descent: given a dataset $\{(\mathbf{x}_i,y_i)\}_{i \leq N}$, the gradient of $L(\mathbf{w})$ at each $(\mathbf{x}_i, y_i)$ could be defined as:
\begin{center}
$\nabla |_{\mathbf{x}_i,y_i}L(\mathbf{w})=-(y_i-\sigma(\mathbf{w}^\top\mathbf{x}_i))\mathbf{x}_i$,
\end{center}
where $\sigma(z)=\frac{1}{1+e^{-z}}$ is the sigmoid function.

\subsection{Secure Multi-party Computation (MPC) and ABY3}
MPC is a set of cryptographic techniques allowing parties to jointly compute a public function over their private inputs. Many different MPC protocols specialized for ML have been introduced under various security models, examples include EzPC \citep{chandran2017ezpc} and SecureML \citep{mohassel2017secureml}. iDASH seeks a solution in the 3-party semi-honest (with honest majority) model, and so far the best protocol satisfying that model is \textbf{ABY3} proposed by \citet{mohassel2018aby}, so we chose ABY3 as the underlying protocol of our solution.
Due to space limitations, we briefly describe ABY3 in Appendix~\ref{TFEcode} and refer the reader to their paper for further details.

%% file: 3_Solution.tex
\section{Solution description}
\label{solution}

\subsection{Optimized logistic regression in MPC}





  
\mypara{Class weight adjustment for imbalanced training set}
The numbers of positive and negative samples in the given dataset are imbalanced, which inflates false positives for the majority class. A common countermeasure is resampling, which either remove samples from the majority class or add more duplicated examples from the minority class. But in MPC, the labels are private so resampling is difficult. Differential privacy is another possible solution, but it will lead to a non-negligible accuracy loss for such a small dataset. So we adopt the simpler yet well-motivated approach of weighting samples in the loss function to place more emphasis on minority classes:
\begin{center}
$C_0 = \frac{N}{2\sum_{i=1}^{N}(1-y_i)}, C_1 = \frac{N}{2\sum_{i=1}^{N}y_i}$
\end{center}
and the gradient becomes: $\nabla L(\mathbf{w})=-(y_i-\sigma(\mathbf{w}^\top\mathbf{x}_i))\mathbf{x}_i\cdot C_{y_i}$. Note that we cannot lookup $C_{y_i}$ directly since $y_i$ is a private input, so instead  we compute it via $C_{y_i} = (C_1 - C_0) \cdot y_i + C_0$.

\mypara{Parallel piecewise approximation of the sigmoid function}
The sigmoid function in LR has to calculate exponents, which is not practical in MPC. The original ABY3 paper uses piecewise polynomials (described in Algorithm~\ref{alg:poly_alg}) to approximately compute sigmoid. In particular, they use a 3-piece approximation, which unfortunately achieves subpar accuracy relative to the non-secure model (See Table~\ref{tbl:3vs5}). Instead, we use a 5-piece approximation as shown in Figure~\ref{3vs5}. 

\begin{minipage}{.52\textwidth}
\begin{algorithm}[H]
\begin{algorithmic}[1]
\caption{Piecewise polynomial function} \label{alg:poly_alg}
\Require 
\item[] Private input: $x$;
\item[] Segmentation points: $\{s_i\}_{0\leq i \leq n}$, where $s_0=-\infty$ and $s_n = \infty$;
\item[] Expressions for $n$ polynomials $\{f_i\}_{0\leq i < n}$; 
\Ensure $f_i(x)$ if $s_i < x \leq s_{i+1}$
\For{$i:=0~\text{to}~n$}             
    \State $p_i \gets (x < s_i)$ 
\EndFor
\For {$i:=0~\text{to}~n-1$}
    \State $b_i \gets \neg p_i \land p_{i+1}$
\EndFor
\State $r \gets \sum_{i=0}^{n-1} b_i \cdot f_i(x)$ 
\end{algorithmic}
\end{algorithm}
\end{minipage} 
\hspace{0.25cm}
\begin{minipage}{.45\textwidth}
\begin{figure}[H]
\includegraphics[width=6cm]{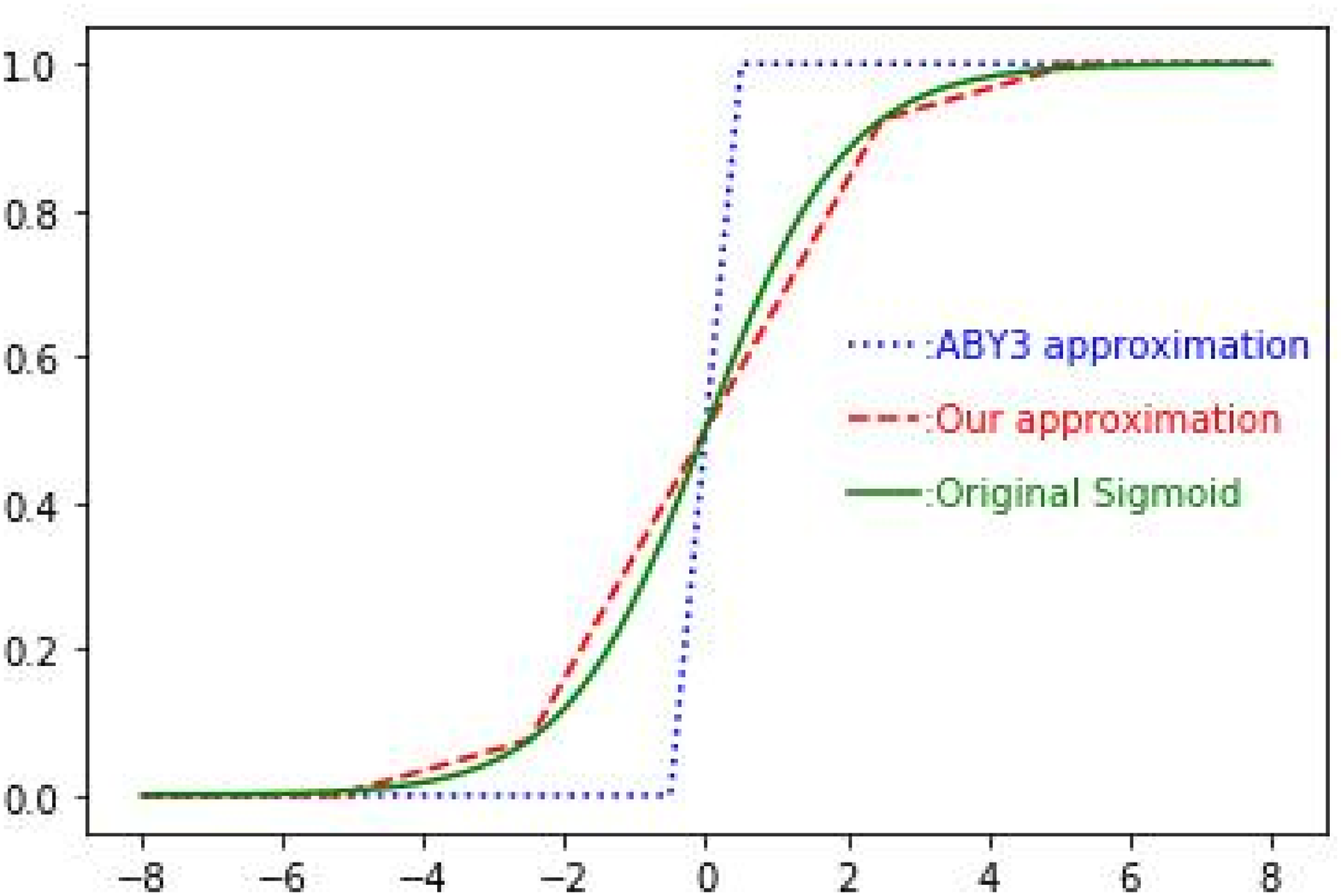}
\caption{Different approximate sigmoid functions. Function bodies are deferred to Appendix A.2 for space limit.}
\label{3vs5}
\end{figure}
\end{minipage}

\subsection{MPC programming in TF Encrypted}

\textbf{TensorFlow}~\citep{TensorFlow16} is a high-performance ML framework with built-in support for executing distributed dataflow graphs across a cluster of servers, and \textbf{TF Encrypted (TFE)} is an open-sourced layer on top of TensorFlow for privacy-preserving computations. TFE has a TensorFlow-like interface and allows ML engineers who are non-experts in cryptography to easily develop encrypted machine learning solutions. For instance, with an existing TFE MPC protocol, implementing secure LR training is just 20 lines of code. An example could be found in Appendix~\ref{TFELR}.

\mypara{Advantages of TF Encrypted \& TensorFlow in MPC programming} 
In TensorFlow, code execution is broken into two phases: graph building and graph execution. The former is typically done through a high-level Python API, while the latter is performed by a C++ runtime that eventually partitions the distributed graph into a set of local operations to be executed by the MPC servers. This separation allows the engine to optimize the computation at several levels.

One concrete benefit of this declarative approach is that parallel execution is determined at the discretion of the runtime and not the programmer. Take Algorithm 1 for example, $n$ comparisons in Step 2 will be run in parallel automatically. 
Although this parallelization could be manually achieved in other frameworks such as~\cite{aby3}, it is made conveniently automatic in TensorFlow.

Another benefit of the distributed graph representation is that it hides the low-level networking operations. Tensors held by one server may seamlessly be used by another, and it is the responsibility of the runtime to insert the required send/recv operations during graph partitioning, and may even choose a communication technology optimized for the runtime environment. In other words, the programmers do not have to deal with complicated socket programming, which is an important component of MPC protocols and easy to get wrong. An example could be found in Appendix~\ref{TFEcode}.

Finally, as a framework for large-scale machine learning, TensorFlow ships with several useful functionalities, including high-performance data pipelines, a high-level language for building machine learning models, and abstractions for clusters of potentially heterogeneous compute devices.

\mypara{ABY3 as a TFE module}
Considering all the advantages above, we chose TF Encrypted as the underlying framework, and implemented the ABY3 protocol on top of it. The code has been contributed to TFE as one of its protocol modules. 

%% file: 4_Results.tex
\section{Results}
\label{results}

We test our implementation using three ecs.g6.2xlarge instances of Alibaba Cloud, each with an 8--core 2.50 GHZ CPU and 32GB of RAM. 

\subsection{Performance of the TFE ABY3 framework}
To evaluate our TFE ABY3 implementation, we compare with \cite{aby3}, a C++ framework written by the authors of ABY3. 
The results in Table~\ref{tbl:benchmark} indicate that we are $1.5-1.8\times$ slower for small-scale data , while $1.1-1.8\times$ faster for large-scale data because of TensorFlow's graph optimizations\footnote{We make these comparisons in localhost because their code lack a WAN test case. The relative difference is expected to disappear in WAN, because latency will dominate any extra computational time.}. 
The reason is that for small-scale data, the higher amount of iterations per second leads to more frequent API calls, so the overhead introduced by the TensorFlow C++ wrappers is non-negligable. On the other hand, for large-scale data, TensorFlow's graph optimizations begin to show advantages from many aspects, such as automatic parallel processing and faster matrix multiplication.

\begin{table}
\caption{Comparison of LR training speed, measured in iterations per second (larger = better). Both frameworks use the 3-piece sigmoid approximation.}
\label{tbl:benchmark}
\begin{center}
\begin{tabular}{llcccc}
\multicolumn{1}{c}{\bf Batchsize}  &\multicolumn{1}{c}{\bf Framework } &\multicolumn{1}{c}{\bf 64 features} &\multicolumn{1}{c}{\bf 1024 features} &\multicolumn{1}{c}{\bf 4096 features} &\multicolumn{1}{c}{\bf 16384 features}
\\
\hline
\multirow{2}*{64} & 
\citet{aby3}   & $1408$  & $704$   &$257$  &$52$      \\ 
~& TFE ABY3    & $824$   & $564$  & $225$  & $\textbf{58}$  \\
\hline
\multirow{2}*{128} & 
\citet{aby3}   & $1265$  &  $421$   & $114$   & $25$      \\ 
~& TFE ABY3   & $754$  & $380$  & $\textbf{124}$  & $\textbf{30}$       \\
\hline
\multirow{2}*{256} & 
\citet{aby3}   & $1068$  &  $209$   & $34$  & $8.5$         \\ 
~& TFE ABY3   & $720$  & $\textbf{211}$  & $\textbf{60}$    & $\textbf{11}$    \\
\end{tabular}
\end{center}
\end{table}

\subsection{Performance of the iDASH solution}

We preprocess the dataset with a feature selection step (see Appendix~\ref{bioinformatics}) and test our model with 10-fold cross-validation. To make the training algorithm generalizable, we use the ``optimal'' adaptive learning rate (See Appendix~\ref{learnrate}) proposed by \citet{bottou2012stochastic}, and set L2 penalty = 1. All the evaluations in this subsection are written using TFE ABY3.

\mypara{Accuracy} We test the balanced accuracy $(\frac{TP}{TP+FN}+\frac{TN}{FP+TN})/2$ using the three Sigmoid functions described in Figure~1\footnote{Only GSE2034 is described here because BC-TCGA is easily separable (accuracy 100$\%$) thus omitted.}. The result in Table~\ref{tbl:3vs5} shows that the 3-piece one is unsatisfactory for this case, while our 5-piece one are nearly as accurate as the cleartext version (about $70\%$).



\mypara{Training time} We test on two cases: a LAN network with 1Gbps bandwidth and sub-milisecond latency, and a WAN network with 100Mbps bandwidth and 25 miliseconds latency (i.e., 50 miliseconds round trip time). The result is shown in Table~\ref{tbl:3vs5}. We can see that the 5-piece Sigmoid is $1.5-1.7\times$ slower in LAN because of more computational cost, but there's no observable difference in WAN because the pieces are evaluated in parallel, thus they trigger the same number of send/recv operations, and the latency dominates the total training time. We emphasize again that this parallelism is automatically achieved by TensorFlow, rather than manually multithread coding.

\begin{table}[tb]
\setlength{\belowcaptionskip}{5pt}
\begin{center}
\caption{Comparison of Sigmoid approximations. Training speed measured in iterations per second. }

\label{tbl:3vs5}
\begin{tabular}{llcccc}
{\multirow{2}{*}{\bf Methods}} &{\multirow{2}{*}{\bf Batchsize}}
&\multicolumn{2}{c}{\bf Training speed }
&\multicolumn{2}{c}{\bf Accuracy}
\\
\cmidrule(lr){3-4} \cmidrule(lr){5-6}
~ & ~
&{\bf  LAN} &{\bf WAN} &{\bf  100 iterations} &{\bf 200 iterations} 
\\
\hline
ABY3's 3-piece Sigmoid   & \multirow{2}{*}{32} & $834$  & $2.09$  & $63.13\%$  & $64.19\%$   \\ 
Our 5-piece Sigmoid    & ~&$496$   & $2.09$  & $66.42\%$  & $67.54\%$   \\
\hline
ABY3's 3-piece Sigmoid   & \multirow{2}{*}{64} & $800$  & $2.08$  & $58.45\%$  & $65.22\%$   \\ 
Our 5-piece Sigmoid    & ~&$475$   & $2.08$  & $67.47\%$  & $69.48\%$   \\
\end{tabular}
\end{center}
\end{table}

%% file: Appendix.tex
\appendix
\section{Appendix}

\subsection{Domain-specific knowledge from bioinformatics}
\label{bioinformatics}

We noticed that the datasets are about breast cancer, and there are already several works that have picked out the most significant genomes (signatures) to predict the clinical outcome of breast cancer \citep{ross2008commercialized}. A few examples are Oncotype DX™ \citep{paik2004multigene}, MammaPrint®\citep{van2002gene}, the Rotterdam Signature \citep{wang2005gene}, and the Invasiveness Gene Signature (IGS) \citep{liu2007prognostic}. After testing multiple signatures on both datasets with cross validation, we found that the Rotterdam Signature of 76 genes (but only 67 of them exist in both datasets) turns out to be the best performing one, boosting the balanced accuracy (with our Sigmoid approximation) from $62$--$66\%$ to $68$--$71\%$. It is worth mentioning that it's almost impossible to pick those signatures using feature selection methods, because they are picked based on lots of extra information that is not listed in the dataset, such as genome functions, medical diagnosis and relapse time.

\mypara{Separating breast cancer subtypes}
The accuracy could be further improved by exploiting deeper domain-specific knowledge: Breast cancer could be divided into several subtypes, and different subtypes are correlated with different genes. As \cite{carey2007triple} describes, breast cancers could be divided into ER+ and ER- subtypes based on ER status, which is correlated with the gene ``ESR1''. Specifically, 16 genes of the Rotterdam Signature are related with
ER-, and the other 60 are related with ER+.

Based on the above fact, a two-model solution could be developed:  Two LR models are trained on the dataset, the first one uses the full Rotterdam Signature as training features, while the second one uses only the 16 ER- genes as training features. When we want to predict on a patient X, we first observe X's gene expression of ``ESR1''. If it's significantly lower than average, it indicates a high chance that X belongs to the ER- subgroup, and the second model will be used to predict. Otherwise the first one will be used.

We did experiments and found that this new method could increase the accuracy significantly (sometimes above $76\%$), and submitted it to the competition. But due to some misunderstandings, this two-model solution was not tested by the organizers, and only the one using the full signature (with accuracy $68$--$70\%$) was tested.

\subsection{Piecewise functions used}
\label{piece}
$\mathtt{ABY3}: \sigma(x)=\begin{cases}0, \quad x\textless -0.5 \cr x+0.5, \quad -0.5\leq x \textless 0.5 \cr 1, \quad x\ge 0.5 \end{cases}$\\

$\mathtt{Ours}: \sigma(x) = 
\begin{cases}
10^{-4}, \quad x \le -5\\
0.02776\cdot x + 0.145, \quad -5 < x \le -2.5\\
0.17\cdot x + 0.5, \quad -2.5 < x \le 2.5\\
0.02776\cdot x + 0.85498, \quad 2.5 < x \le 5\\
1 - 10^{-4}, \quad x > 5
\end{cases}$

\subsection{A little more details of ABY3 in TensorFlow}
\label{TFEcode}
\mypara{Replicated secret sharing} In ABY3, each private
data $x\in Z_{2^k}$ ($k$ is the length of bits we used to represent a number, e.g. $k=64$) is secret-shared to three parties $P_0,P_1,P_2$, using one of the following three kinds of techniques:
\begin{itemize}
    \item \textbf{A}rithmetic sharing: Sample
three random values $x_0, x_1, x_2 \in Z_{2^k}$ , such that $x =
x_0 + x_1 + x_2$. Each party $P_i$ holds $x_i$ and $x_{i+1\mod 3}$. We call these shares as ``A-shares'' of $x$, denoted as $[\![x]\!]$.

    \item \textbf{B}oolean sharing: Sample
three random values $x_0, x_1, x_2 \in Z_{2^k}$ , such that $x = x_0 \oplus x_1 \oplus x_2$. Each party $P_i$ holds $x_i$ and $x_{i+1\mod 3}$. We call these shares as ``B-shares'' of $x$, denoted as $<\!<x>\!>$.

    \item \textbf{Y}ao sharing: Yao sharing is used for Garbled Circuit (GC). Unfortunately we found that GC is inefficient in TensorFlow, so we decided to use only A/B shares here. Y shares are not discussed in this paper and left for our future work.
\end{itemize}

\mypara{Addition and substraction}
Given two A-shared data $[\![x]\!]$ and $[\![y]\!]$, it's easy to see that addition (and substraction) could be done locally: $[\![x+y]\!] = [\![x]\!]+[\![y]\!] $

\mypara{Multiplication}
To multiply two shared values $[\![x]\!]$ and $[\![y]\!]$, Party $P_i$ locally computes $z_i=x_iy_i + x_iy_{i+1\mod 3} + x_{i+1\mod 3}y_i$\footnote{The zero sharings are omitted in this section for clearer understanding}, and send $z_i$ to party $P_{i-1\mod 3}$. We can see that $[\![z]\!]$=$[\![xy]\!]$ . Similar method works for matrix multiplication, and an example code of multiplication could be found in Appendix~\ref{TFEcode}.

Other operations that we have implemented include:
\begin{itemize}
 \item Sharing conversion between A-shares and B-shares.
 \item Bit extraction, MSB extraction, and Comparison.
 \item Piecewise polynomial evaluation, Sigmoid approximation.
\end{itemize}

We refer the reader to the original paper \citep{mohassel2018aby} for further details of these operations.

\mypara{ABY3 in TensorFlow} The Tensor data structure, which is the output of some computation primitive, will stay only on the server that executes the primitive, unless it is explicitly instructed to be used in another server. This gives an intuition about the natural integration between ABY3 and TensorFlow:
each pair of shares $(x_i; x_{i+1\%3})$ will be held by
its corresponding party $i$, and most computation primitives are
defined purely on each server’s local shares; when a server
uses a share held by another server, it is equivalent to a network send/recv operation between the two servers. 

Take the following code for example, the script $z[1][1] = z2$ in the re-sharing step triggers $z2$ to be send from $servers[2]$ to $servers[1]$, and no network programming is needed.


\begin{lstlisting}[language=python]
def _matmul_private_private(prot, x, y):
  assert isinstance(x, ABY3PrivateTensor), type(x)
  assert isinstance(y, ABY3PrivateTensor), type(y)

  x_shares = x.unwrapped
  y_shares = y.unwrapped

  # Tensorflow supports matmul for more than 2 dimensions,
  # with the inner-most 2 dimensions specifying the 2-D matrix multiplication
  result_shape = tf.TensorShape((*x.shape[:-1], y.shape[-1]))

  z = [[None, None], [None, None], [None, None]]
  with tf.name_scope("matmul"):
    a0, a1, a2 = prot._gen_zero_sharing(result_shape)

    with tf.device(prot.servers[0].device_name):
      z0 = x_shares[0][0].matmul(y_shares[0][0]) \
           + x_shares[0][0].matmul(y_shares[0][1]) \
           + x_shares[0][1].matmul(y_shares[0][0]) \
           + a0

    with tf.device(prot.servers[1].device_name):
      z1 = x_shares[1][0].matmul(y_shares[1][0]) \
           + x_shares[1][0].matmul(y_shares[1][1]) \
           + x_shares[1][1].matmul(y_shares[1][0]) \
           + a1

    with tf.device(prot.servers[2].device_name):
      z2 = x_shares[2][0].matmul(y_shares[2][0]) \
           + x_shares[2][0].matmul(y_shares[2][1]) \
           + x_shares[2][1].matmul(y_shares[2][0]) \
           + a2
    # Re-sharing
    with tf.device(prot.servers[0].device_name):
      z[0][0] = z0
      z[0][1] = z1
    with tf.device(prot.servers[1].device_name):
      z[1][0] = z1
      z[1][1] = z2
    with tf.device(prot.servers[2].device_name):
      z[2][0] = z2
      z[2][1] = z0

    z = ABY3PrivateTensor(prot, z, x.is_scaled or y.is_scaled, x.share_type)
    z = prot.truncate(z) if x.is_scaled and y.is_scaled else z
    return z
\end{lstlisting}

\subsection{20 lines of code for LR in TFE}
\label{TFELR}

Once the underlying primitives (namely addition, substraction, multiplication, sigmoid, matrix multiplication) were done, an ML engineer with little background in cryptography could develop privacy-preserving ML programs easily. As is shown in the example, logistic regression training in MPC could be done in 20 lines of code, with few differences compared to common TensorFlow code.

\begin{lstlisting}[language=python]
def logistic_regression():
    prot = ABY3()
    tfe.set_protocol(prot)

    # define inputs
    x = tfe.define_private_variable(x_raw, name="x")
    y = tfe.define_private_variable(y_raw, name="y")
    # define initial weights
    w = tfe.define_private_variable(tf.random_uniform([10, 1], -0.01, 0.01),name="w")
    learning_rate = 0.01

    with tf.name_scope("forward"):
      out = tfe.matmul(x, w) + b
      y_hat = tfe.sigmoid(out)

    with tf.name_scope("loss-grad"):
      dy = y_hat - y
      
    batch_size = x.shape.as_list()[0]
    with tf.name_scope("backward"):
      dw = tfe.matmul(tfe.transpose(x), dy) / batch_size
      assign_ops = [tfe.assign(w, w - dw * learning_rate)]

    with tfe.Session() as sess:
      # initialize variables
      sess.run(tfe.global_variables_initializer())
      for i in range(1):
        sess.run(assign_ops)
\end{lstlisting}

\subsection{The ``optimal'' learning rate}
\label{learnrate}
In Section 5.2 of \cite{bottou2012stochastic}, they proposed the following adaptive learning rate:
\begin{equation}
\eta = \eta_0 / (1 + \lambda \eta_0 t)
\end{equation}
, where $\eta_0$ is a heuristic initial value, $\lambda$ is the decreasing factor, and $t$ means it's the $t^{th}$ iteration.

This formula has been proved to be effective in practise, and is used in the SGDClassifier of Scikit-learn when we set $learning\_rate='optimal'$.

For our implementation, we set
$\eta = 1 / (1.2 + t)$.